\documentclass[aps,prb,twocolumn,superscriptaddress,noshowpacs,a4paper,10pt]{revtex4-1}
\usepackage{color}
\usepackage{amssymb}
\usepackage{graphicx}
\usepackage[pdftex]{hyperref}
\hypersetup{colorlinks=true,citecolor=blue,linkcolor=blue,urlcolor=blue}
\usepackage[all]{hypcap}

\begin{document}

\author{\firstname{Roberto} \surname{Guerra}}
\author{\firstname{Stefano} \surname{Ossicini}}
\affiliation{Dipartimento di Scienze e Metodi dell'Ingegneria and Centro Interdipartimentale En\&Tech, Universit\`a di Modena e Reggio Emilia, via Amendola 2 Pad.\ Morselli, I-42122 Reggio Emilia, Italy.}

\title{ The Role of Strain in Interacting Silicon Nanoclusters }

\begin{abstract}\textbf{
The possibility of controlling the optical transition probability between neighbouring silicon nanoclusters (Si-NCs) constitutes nowadays an attractive prospect in nanophotonics and photovoltaics. In this work, by means of theoretical \textit{ab initio} calculations we investigate the effect of strain on the opto-electronic properties of Si-NCs pairs. We consider two sources of strain: the strain induced by an embedding SiO$_2$ matrix, and the strain generated by mutual NC-NC forces occurring at small distances. Independently on its source, we observe a fundamental impact of the strain on the orbitals localization and, as a consequence, on the transition probability between energy states, belonging or not to the same NC. The resulting picture allots to the structural strain a fundamental role in the NC-NC interaction mechanisms, suggesting the possibility of enabling a strain-controlled response in Si-NC ensambles.
}\end{abstract}

\maketitle

\section{Introduction}\label{sec.intro}
\noindent Highly promising results have been obtained in the latest years from Si nanoclusters (NCs) in many fields, among which photonics,\cite{nature_pavesi,bourianoff} non volatile memories,\cite{tiwari,yao} and biological applications,\cite{erogbogbo} while trial photovoltaics devices are under investigation.\cite{osso_nanoreslett} In all the cases, the main advantage of this system comes from the chance of tuning the optical response by changing the NC size and other structural characteristics.
In addition, evidence of an interaction mechanism operating between NCs has been frequently reported,\cite{heitmann_zacharias,linnros,glover_meldrum} sometimes indicated as an active process for optical emission,\cite{iwayama} and sometimes even exploited as a probing technique.\cite{schneibner}. Moreover a mechanism based on the transformation of high-energy photons into low-energy electron-hole pairs, via multiple exciton generation (MEG), localized on neighbouring Si nanocystals has been proposed \cite{timmerman,trinh} and elucidated \cite{ivan} as suitable route to minimize solar cell loss factors. 
\\Nevertheless, while the role of the size, shape, interface configuration, embedding medium, on the otpoelectronic properties of single NCs has been extensively investigated \cite{ossicini,degoli,dalnegro,guerrasin,loeper}, the study of the effects of NC-NC interplay has received less attention. In the simplest picture, when the separation between the NCs lowers under a certain limit, the wave functions overlap with the neighbour ones, promoting the tunneling process and limiting the quantum-confinement (QC) effect.
\\From the experimental point of view it is difficult to control the size and the distance between the nanocrystals, even if progresses have been made in these last years \cite{zacharias,heitmann_zacharias,lockwood_meldrum,godefroo}. Theoretically the impact of NC-NC separation on their optolectronic properties has been only recently considered \cite{allan_delerue,galli,seino,ivan}. Allan and Delerue \cite{allan_delerue} have studied the energy transfer between Si nanocrystals showing that the transfer is possible only when the dots are almost in close contact. 
Seino et al.\cite{seino} have studied, using ab-initio density functional theory (DFT) based methods, the impact of NC size and NC-NC separation on the electronic properties and carrier transport for Si NCs embedded in a SiO$_2$ matrix. Their main conclusion was that, at small separation (0.2\,nm), the energy levels of the NC are broadened to minibands due to wave-function overlap, thus enabling electron trasport. In the calculations they consider a three-dimensional arrangement of nearly spherical Si NCs in a simple cubic lattice with a single NC per unit cell and the atomic geometry is optimized. The variation in the distance between neighbouring NCs is obtained by changing the dimension of the embedding matrix, therefore all the NCs in the calculations are equivalent.
A similar approach has been used by Gali et al.\cite{galli} for H-terminated Si NCs. They consider two different arrangements for the Si NCs in their supercell, a configuration where the distance between neighbouring NCs is the same in each directions and another configuration where two neighbouring NCs are closer to each other only in one direction. Looking at the results for NC absorption they concluded that the absorption clearly increases in all energy ranges as the NCs approach each other.
Again, Govoni et al.\cite{ivan} have studied energy transfer, charge transfer and carrier multiplication (CM) effects, adopting a fully ab-initio scheme within DFT, in both isolated and interacting H-passivated Si-NCs. A side-by-side comparison of the calculated electron-and hole initiated CM lifetimes, demonstrated the existence of a lifetime hierarchy, thus explaining the impact of NC-NC interaction on CM dynamics.
Finally, Lin et al.\cite{lusk} have made use of many-body Green function analysis and first-order perturbation theory to quantify the influence of size, surface reconstruction, and surface treatment on exciton transport between small Si NCs. Their analysis shows that QC causes small ($\sim$1\,nm) Si NCs to exhibit exciton transport efficiencies far exceeding that of their larger counterparts for the same center-to-center separation. They also find that surface reconstruction significantly influences the absorption cross section and leads to a large reduction in both transport rate and efficiency.
\\The influence of strain on the optoelectronic properties of semiconductor materials becomed in the last years an hot topic, expecially regarding
possible device applications. Regarding bulk Si, it has recently demonstrated through a combined experimental-theoretical effort the possibility of using strain to induce second harmonic generation, otherwise prohibited for symmetry reason.\cite{natmat_osso} As far as nanostructures, the main interest has regarded Si, Ge and SiGe nanowires, where DFT calculations have demonstrate their ability, in perspective, to guide the syntheses of nanowires of controlled  shape and geometry for different electronic applications.\cite{nduwimana}
\begin{figure*}[t!]
  \centering
  \includegraphics[draft=false,width=7.5cm]{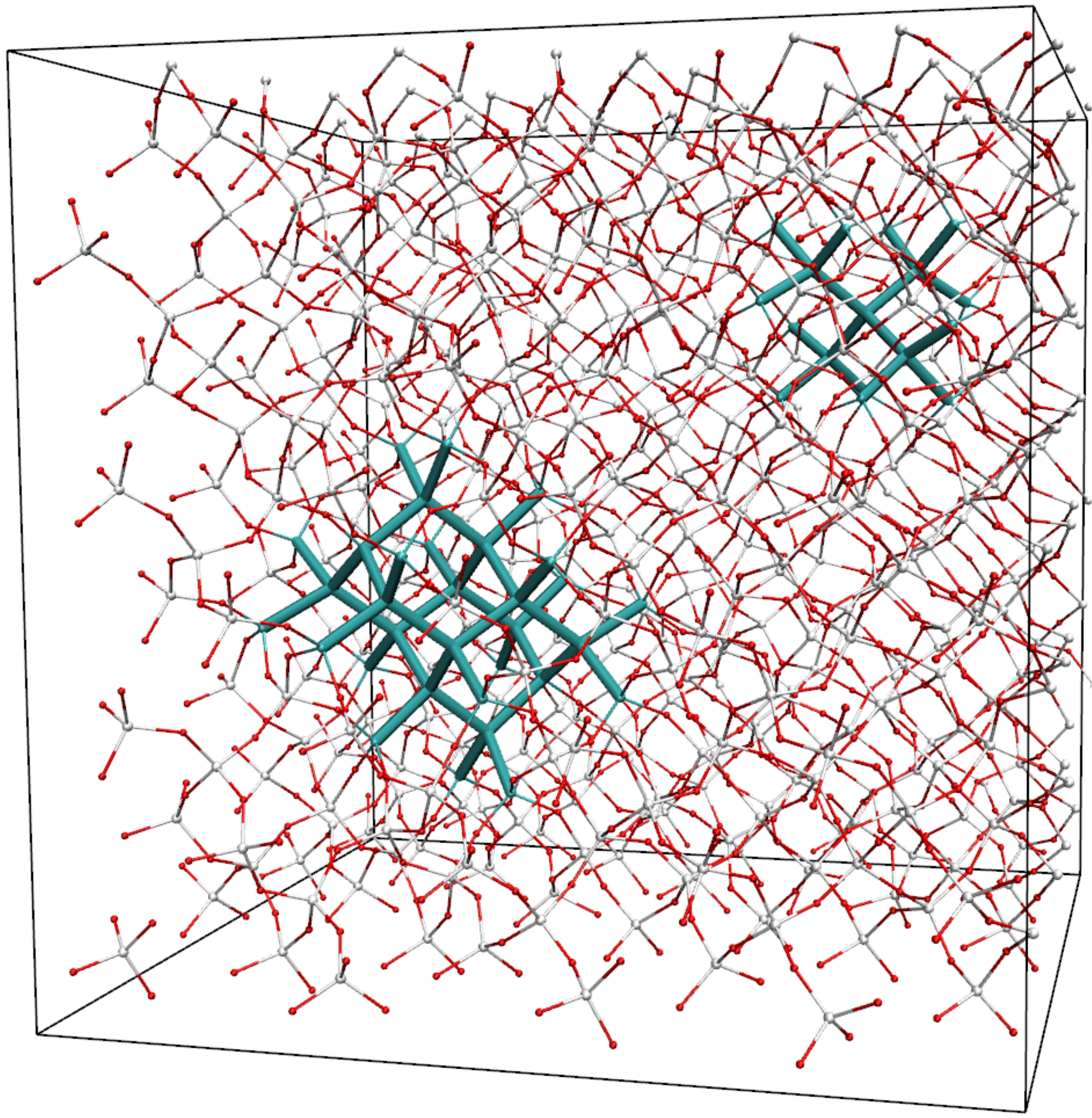}~~~
  \includegraphics[draft=false,width=7.5cm]{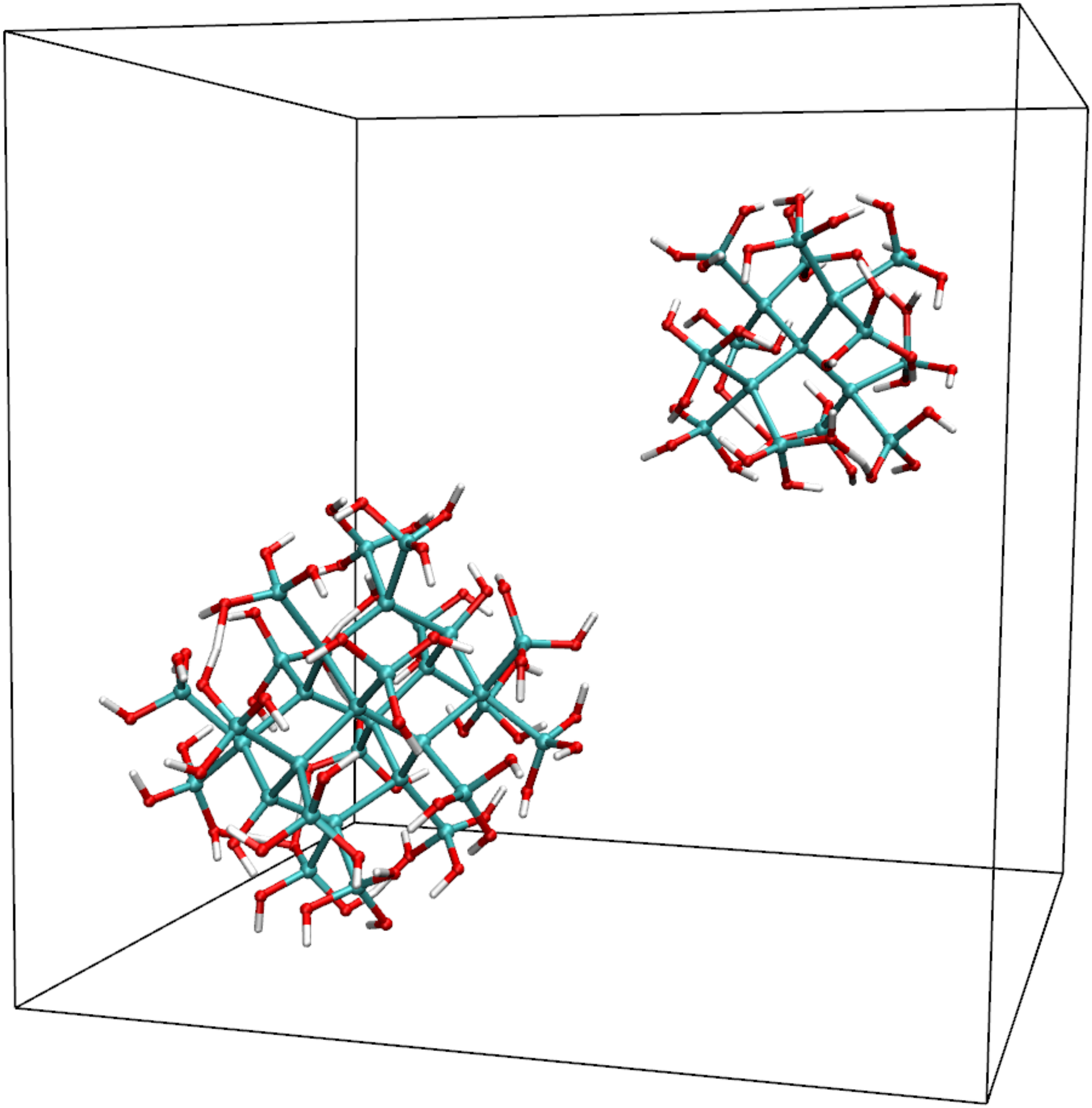}
  \caption{\small (color online) Si$_{17}$+Si$_{32}$ SiO$_2$-embedded (left) and de-embedded (right) NCs pair at a surface-to-surface distance $d$\,=\,0.8\,nm. Si, O, and H atoms are represented in cyan, red, and white, respectively. Si atoms of the SiO$_2$ are represented in white for clarity. The simulation box, displayed by black lines, has side of 2.78\,nm (left) and 2.60\,nm (right). }\label{fig.dots}
\end{figure*}
\\In the present work we focus on the role of strain on the NCs orbital localization, considering both the strain induced by an embedding SiO$_2$ matrix and the strain due to the lowering of the distance between the NCs. In our calculations we consider, in all cases, the presence of two Si NCs in the unit cell, thus we can vary their distance in one direction, while the distance between their periodic images is dictated by the choice of the supercell. We demonstrate that at NC-NC separation lower than a certain threshold strain-inducing forces mutually acting on the NCs emerge, playing an important role in the global response. This aspect should be worth of consideration when dealing with realistic NC ensambles, as for the case of colloidal Si NCs samples\cite{beard,fujii} like as for embedded ones.\cite{kusova,zacharias,heitmann_zacharias}

\subsection*{Structures and Methods}\label{sec.struct-methods}
\noindent In this work we consider a pair of Si-NCs with diameter of about 1\,nm, the Si$_{17}$ and the Si$_{32}$ (in Si$_N$, N is the number of Si atoms forming the nanocrystal), generated by removing the oxygens from two spherical regions of a 4x4x4 betacristobalite-SiO$_2$ sample formed by 1536 Si/O atoms (see Fig.~\ref{fig.dots}, left panel). After a ionic relaxation, the resulting system is formed by the two NCs embedded in the same SiO$_2$ sample and subject to a strain, especially at the Si/SiO$_2$ interface, due to the difference in the lattice spacing of Si and SiO$_2$.\cite{PRB1,PRB2} Concerning the NC-NC separation we have considered two different distances $d$ between the two NCs, $d$\,=\,0.8\,nm and $d$\,=\,0.2\,nm, $d$ indicating the minimum distance between the atoms of one NC and that of the neighbour. 
In order to distinguish the strain induced by the matrix from that induced by the mutual forces between the NCs we get rid of the SiO$_2$ matrix by de-embeddeding the NCs together with the first shell of interface oxygens (in order to mimick the presence of the SiO$_2$\cite{PRB1,PRB2}), preserving the strained geometry of the NCs and by hydrogenating all the dangling bonds in order to avoid the presence of electron states in the band gap. The resulting (de-embedded) system is shown in Fig.~\ref{fig.dots}, right panel. For better understanding the role of NC-NC interaction with respect to that of matrix-induced strain we have also considered the case of two de-embedded NCs placed at a distance of $d$\,=\,0.2\,nm but possessing the original geometry of the two NCs relaxed at $d$\,=\,0.8\,nm.
\\The so-obtained {\it freestanding} NCs have, next, been initially relaxed in separate simulation boxes in order to completely remove the strain (detached relaxation). Then we have considered the case of a rigid placement of the NCs in the same simulation box, and the case of a further structural relaxation that comprise the effects of mutual NC-NC interaction forces (conjoined relaxation). This time the NCs are placed at $d$ of 0.8\,nm, 0.5\,nm, and 0.2\,nm. All the calculations have been performed for the Si$_{32}$+Si$_{17}$ and for the Si$_{32}$+Si$_{32}$ NCs pairs. However, since we have obtained equivalent results for the two cases, for the sake of simplicity we will consider only the first case in the following.
\\Structural, electronic and optical properties have been obtained by full \textit{ab-initio} calculations in the framework of density functional theory (DFT) using the {\small ESPRESSO} package.\cite{espresso} Calculations have been performed using norm-conserving pseudopotentials within the local-density approximation (LDA). An energy cutoff of 60 Ry on the plane-wave basis set has been considered after an apposite convergency test. The optical properties have been calculated within the random-phase approximation (RPA) using dipole matrix elements.\cite{onida_RevModPhys}
For the freestading NCs, in all the calculations we have omitted the vacuum states, i.e. the conduction states of energy equal or above the vacuum energy $E_{vac}$. An estimate of $E_{vac}$ can be obtained by properly aligning the eigenvalues after applying the Makov-Payne correction to the total energy.\cite{makov-payne} In alternative, the vacuum states are identifyable by an inverse-participation-ratio (IPR, see Appendix B) well-below a certain threshold. For each system we have evaluated $E_{vac}$ by a crosscheck of both the above methods.

\section{Electronic and Optical Response}\label{sec.opto-electronic_response}
\noindent At first we discuss the results for the two Si NCs embedded in the SiO$_2$ matrix. In Ref.~\onlinecite{PRB2} we were able to distinguish between the properties that depend only on the NC from those that are instead influenced by the presence of the matrix showing that a single de-embedded NC is able to reproduce very well the absorption spectrum of the full Si/SiO2 system in the energy region up to 7 eV, which is indeed associated to the NC+interface contribution. Instead, the removal of the strain (detached NC) produces an enlargement of the HOMO-LUMO gap and a consequent blue-shift in the absorption spectra in this region (see Table 1 and Figure 2 of Ref.~\onlinecite{PRB2}).
These results were confirmed by Kusova et al.~\cite{kusova} that have collected a large number of experimental data from various sources to demonstrate that free-standing oxide-passivated silicon nanocrystals exhibit considerably blueshifted emission compared to those prepared as matrix-embedded ones of the same size. This effect arise from compressive strain, exerted on the nanocrystals by the matrix, which plays an important role in the light-emission process.
\begin{figure}[b!]
  \centering
  \includegraphics[draft=false,height=4.0cm]{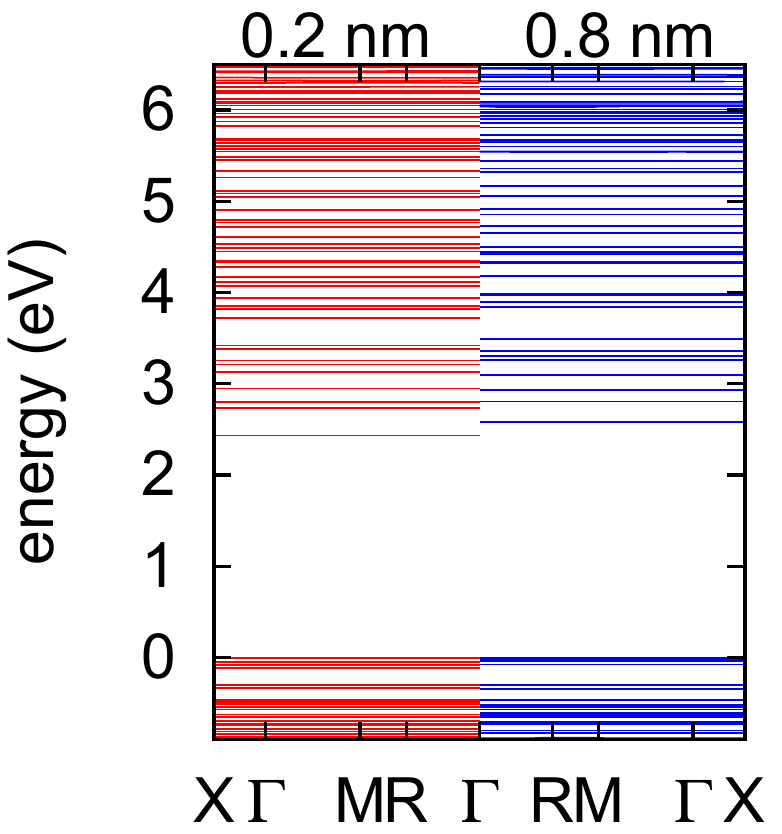}~
  \includegraphics[draft=false,height=4.0cm]{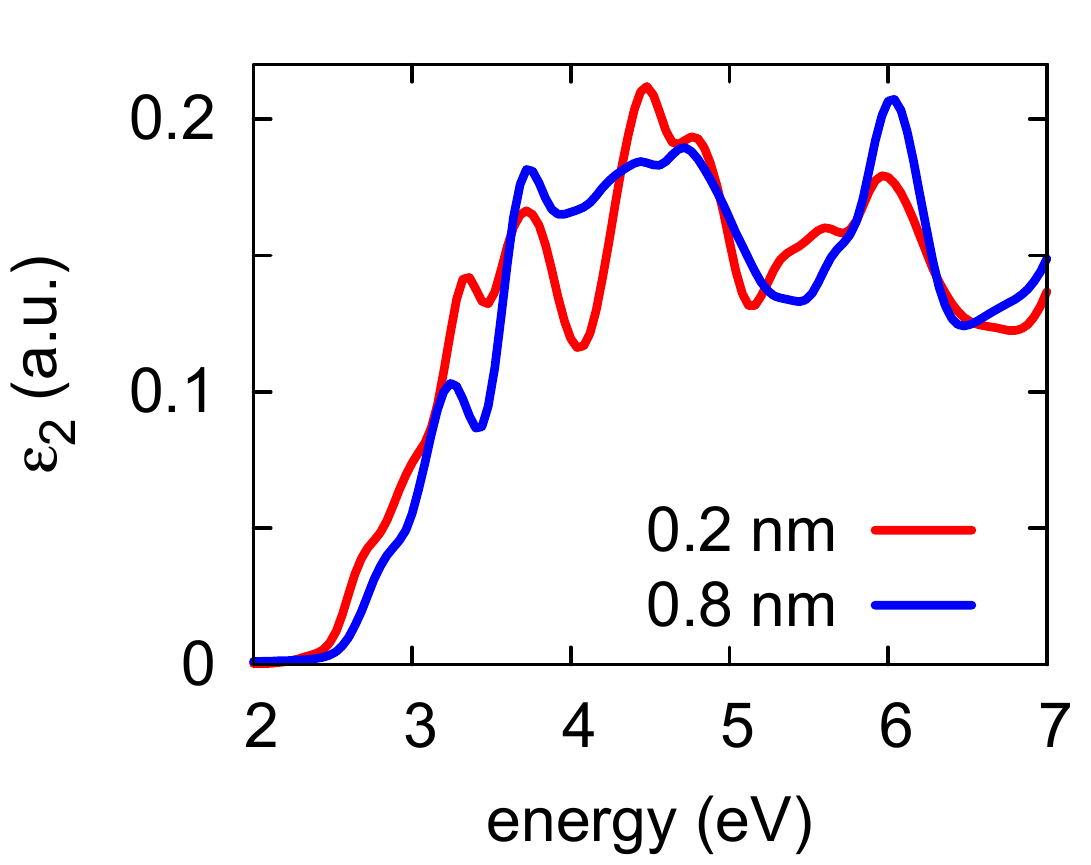}
  \caption{\small (color online) Band structure (left panel) and imaginary part of the dielectric function (right panel) of the Si$_{32}$+Si$_{17}$ NCs pair embedded in SiO$_2$ at $d$\,=\,0.2\,nm and $d$\,=\,0.8\,nm. In the left panel the zero energy corresponds to the top of the valence band. }\label{fig.bande-eps_emb}
\end{figure}

Here, in Fig.~\ref{fig.bande-eps_emb} we report our calculated band structures and  absorption spectra (represented by the imaginary part of the dielectric function) of the embedded NCs pair placed at $d$ of 0.2\,nm and 0.8\,nm. Concerning the bands we note that all the states in the shown energy window are strongly confined, presenting no visible dispersion in energy. This result is at variance with the outcome of Seino et al.\cite{seino} that for a distance between the NCs of $d$\,=\,0.2\,nm observed the formation of minibands due to the interaction between the NCs. The reason of this discrepancy is, in our opinion, due to the fact that Seino et al.\cite{seino} do not consider two NCs embedded in the SiO$_2$ matrix at different distances, but used, instead, a reduction of the matrix thickness in order to simulate two NCs palced at short distance. At the end, in their calculation, one is in presence of a large collection of vicinal Si NCs (due to the presence of images of the NCs in the supercell 
calculation), 
thus their calculation is more suitable for the discussion of the formation of a quantum dots solid, than for elucidating the interaction between single NCs.   
\\Still looking at Fig.~\ref{fig.bande-eps_emb} we note that the energy gap reduces with the distance, consistently with the limit case of connected NCs (i.e. corresponding to a single large cluster). Regarding the absorption spectrum we observe a change of the profile with the distance, while the total integrated absorption is preserved. The latter statement may suggest that no two-sites (NC$_A$--NC$_B$) optical transition contribute to the absorption at $d$\,=\,0.2\,nm, or that the new two-sites transitions arise to the detriment of the one-site (NC$_A$--NC$_A$) ones. This aspect will be clarified in the following.

\begin{figure}[b!]
  \centering 
  \includegraphics[draft=false,width=\columnwidth]{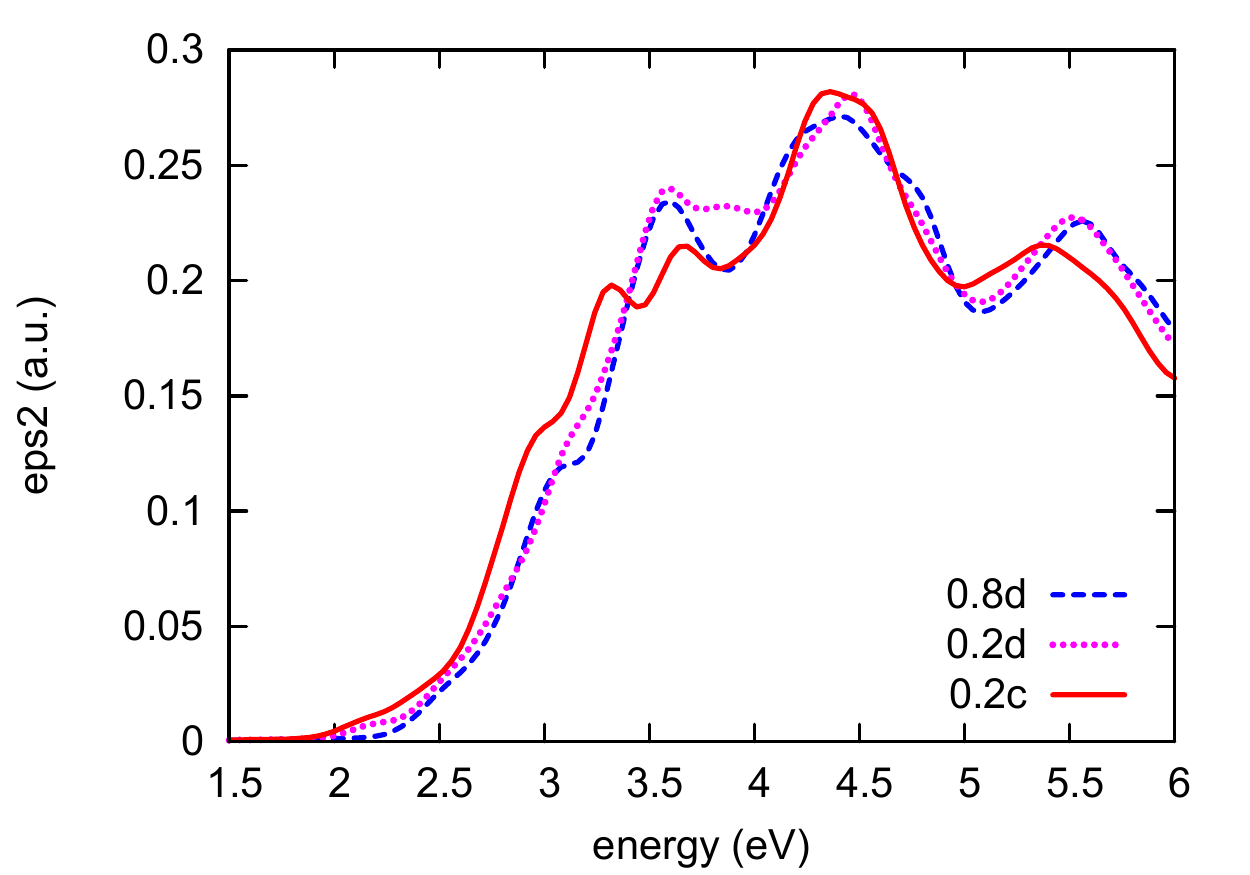}
  \caption{\small (color online) Imaginary part of the dielectric function for the Si$_{32}$+Si$_{17}$ NCs pair relaxed in the SiO$_2$ matrix and then de-embedded. The curves report values for systems relaxed at $d$\,=\,0.8\,nm (0.8d), relaxed at $d$\,=\,0.8\,nm and then rigidly placed at $d$\,=\,0.2\,nm (0.2d), and relaxed at $d$\,=\,0.2\,nm (0.2c). }\label{fig.eps2_de-embedded}
\end{figure}

In order to discern between the role of NC-NC distance and that of strain, we report in Fig.~\ref{fig.eps2_de-embedded} the absorption spectra of the NCs pair de-embedded from the SiO$_2$ matrix, to be compared with those of Fig.~\ref{fig.bande-eps_emb}, right panel. As expected,\cite{PRB1,PRB2} we note a clear resemblance of the spectra with their embedded counterparts. In addition, the embedded NCs pair relaxed at $d$\,=\,0.8\,nm has been de-embedded and rigidly placed at $d$\,=\,0.2\,nm, producing a spectrum (dotted curve) very similar to that of the same pair placed at $d$\,=\,0.8\,nm (dashed curve). The latter result indicates that the strong strain of embedded NCs plays a fundamental role on the final absorption, ruling over the sole variation of $d$ for closely neighbouring NCs.

\begin{figure}[b!]
  \centering
  \includegraphics[draft=false,width=\columnwidth]{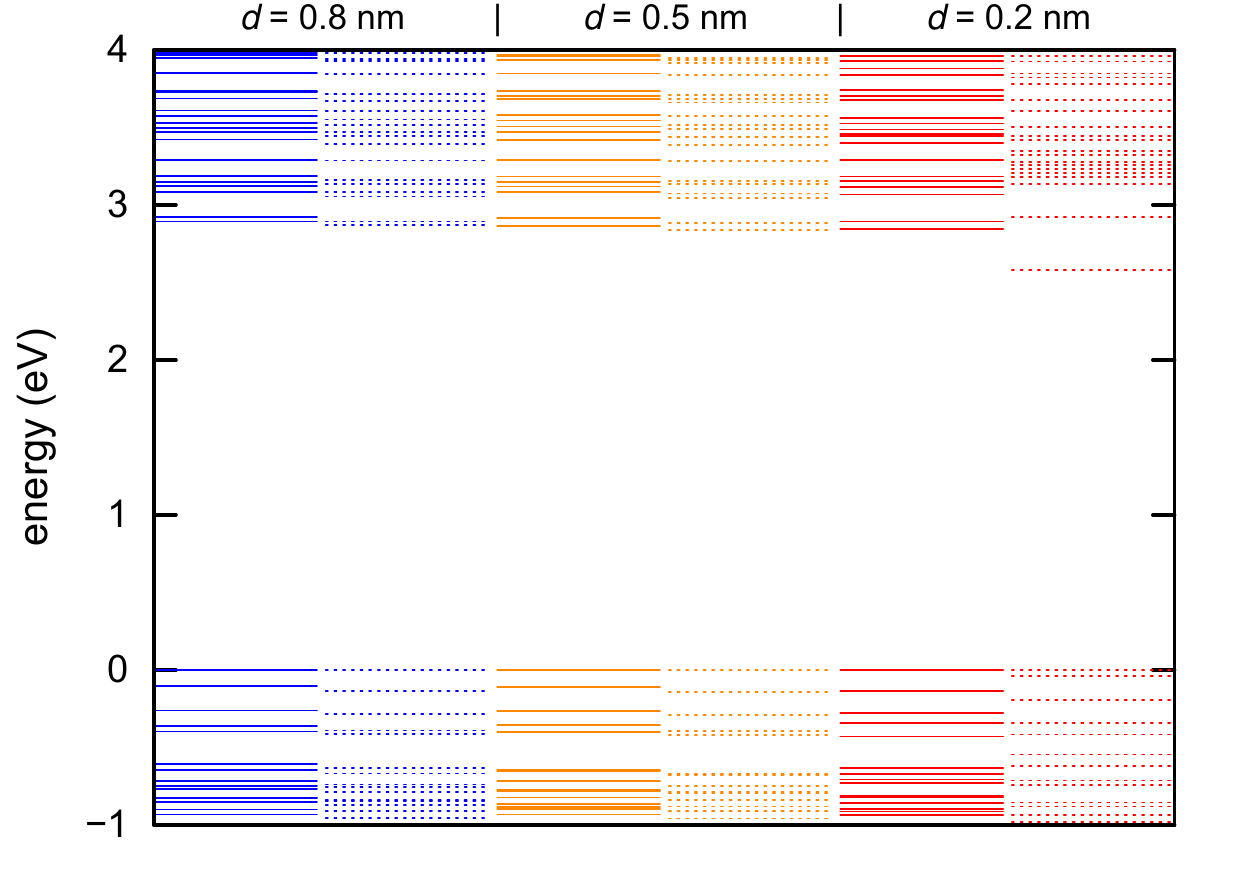}
  \caption{\small (color online) Valence and conduction band-edge energy levels of the Si$_{32}$+Si$_{17}$ freestanding NCs placed at distance $d$\,=\,0.8\,nm (blue), $d$\,=\,0.5\,nm (orange), and $d$\,=\,0.2\,nm (red), for detachedly (solid lines) or conjointly (dotted lines) relaxed structures. The HOMO level marks the zero of energy.}\label{fig.bande}
\end{figure}

\begin{figure}[b!]
  \centering 
  \includegraphics[draft=false,width=\columnwidth]{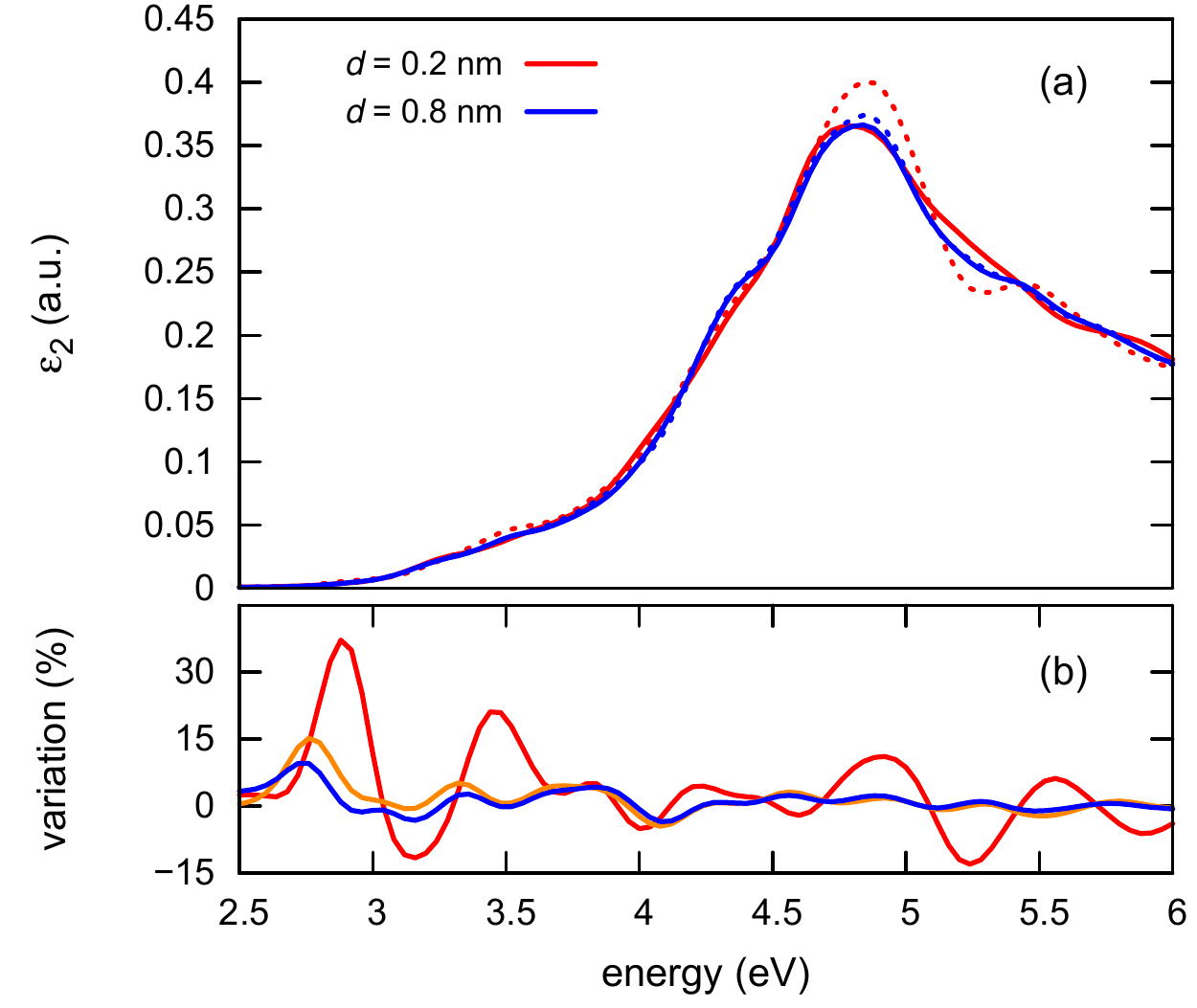}
  \caption{\small (color online) (a) Imaginary part of the dielectric function for the Si$_{32}$+Si$_{17}$ freestanding NCs pair detachedly (solid lines) or conjointly (dotted lines) relaxed in vacuum at distances $d$\,=\,0.8\,nm (blue) and $d$\,=\,0.2\,nm (red). (b) The relative variation between $\varepsilon_2$ of detachedly and conjointly relaxed structures at the same distances and at $d$\,=\,0.5\,nm (orange).}\label{fig.eps2}
\end{figure}

\begin{figure}[b!]
  \centering 
  \includegraphics[draft=false,width=\columnwidth]{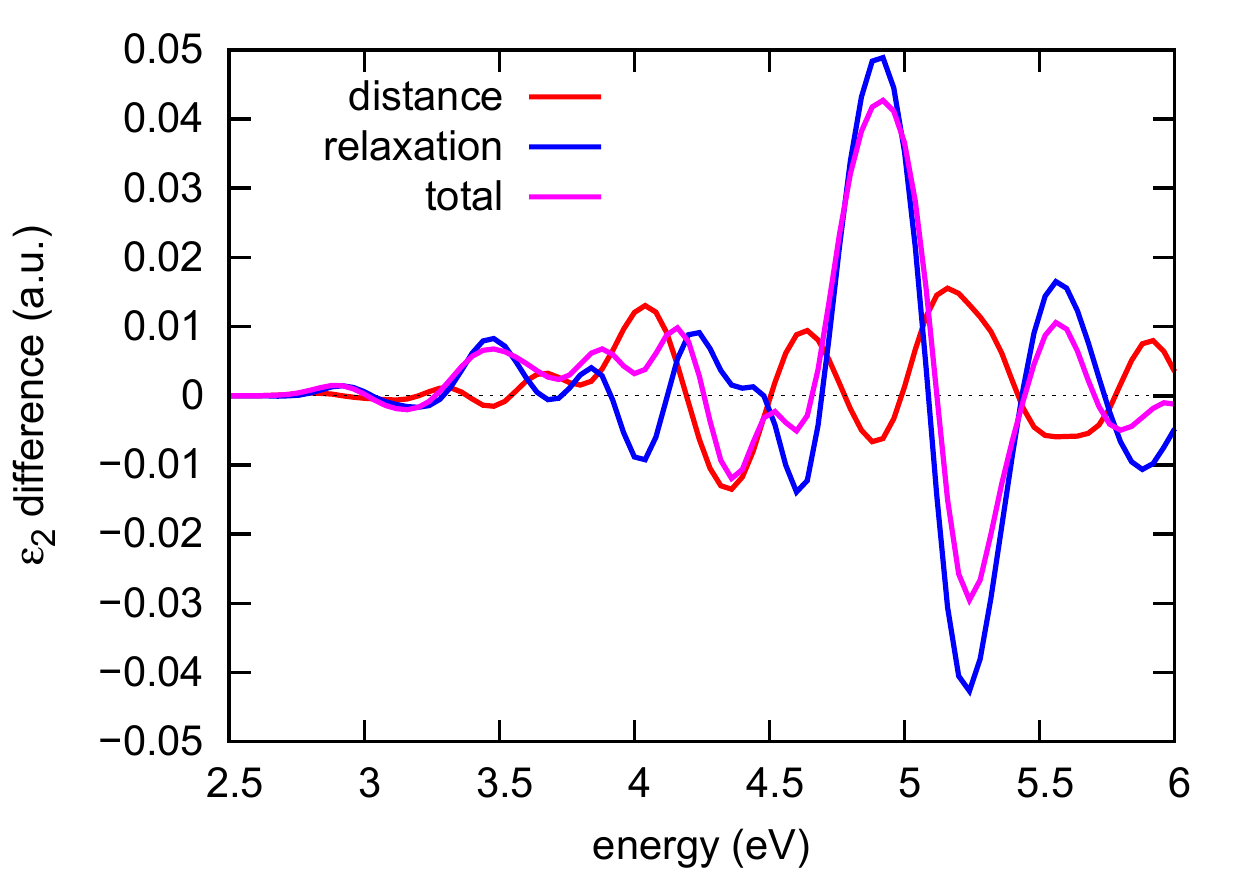}
  \caption{\small (color online) $\varepsilon_2$ variation for the freestanding Si$_{32}$+Si$_{17}$ NCs pair in case of rigid displacement from $d$\,=\,0.8\,nm to $d$\,=\,0.2\,nm with NC geometry detachedly relaxed (red line), of the sole conjoint relaxation at $d$\,=\,0.2\,nm starting from the detachedly relaxed geometry (blue line), and the sum of the two (purple line).}\label{fig.eps2_dist-relax}
\end{figure}

Next we compare the energy levels of the Si$_{32}$+Si$_{17}$ freestanding NCs pair, detachedly or conjointly relaxed, as a function of $d$ (see Fig.~\ref{fig.bande}). First of all we note, as expected for the detachedly relaxed structures, an enlargment of the gap, with respect to the embedded and de-embedded cases, due to the strain relaxation; second we note that at $d$\,=\,0.5\,nm and $d$\,=\,0.8\,nm the effect of the mutual NC-NC forces is negligible, producing small variations of the energy levels for the conjointly relaxed structures. Furthermore, at all $d$ a very weak  variation of the energy levels is observed for the rigidly placed (detached) structures, in agreement with other works.\cite{galli}
Instead, at $d$\,=\,0.2\,nm we observe a reduction of the HOMO-LUMO band-gap (HOMO\,=\,Highest Occupied Molecular Orbital; LUMO\,=\,Lowest Unoccupied Molecular Orbital), of about 10\% when the structures are let free to move. This result clearly indicates a relevant influence of the structural rearrangment induced by the mutual forces over the electronic configuration of the NCs. In this case the reduction of the gap may be addressed to the strain induced by the mutual NC-NC interaction forces. It is interesting to note that $d$\,=\,0.2\,nm is also the separation at which the energy levels of the NCs broaden to minibands for Si NCs embedded in a SiO$_2$ matrix \cite{seino}, making this range of NC-NC distances a particular interesting one. 
\\In order to investigate the effect of the structural relaxation on the optical properties of NC ensambles, we report in Fig.~\ref{fig.eps2}a the absorption spectra  corresponding to the band structures depicted in Fig.~\ref{fig.bande}. From Fig.~\ref{fig.eps2}a we observe that for detachedly relaxed structures there are only small changes in the spectra, whereas  a significant variation of the absorption spectrum due to the structural relaxation is visible on going from $d$\,=\,0.8\,nm to $d$\,=\,0.2\,nm. This variation can go up to 37\% as showed in  Fig.~\ref{fig.eps2}b.
From the same Figure we note that the structural rearrangements show only minor  effect at $d$\,=\,0.5\,nm, making clear that such mechanism is especially relevant in densely packed NC ensambles. Besides, we expect that the threshold $d$ value will depend on the NC size, shape, and passivation type, and cannot therefore hold in general. For example as showed before, in the presence of an embedding matrix, the possibility of NC's structural rearrangement as consequence of their decreasing distance are much limited. This latter aspect will be further discussed in the following.
\\Still from Fig.~\ref{fig.eps2}a we note that, as the variation of $d$, like as the structural rearrangement connected to it, have some role on the NCs pair absorption. It is therefore interesting to compare the effect on the total absorption of the pure variation of the distance (fixed structures) with that of the pure relaxation of the structures (fixed distance). To our knowledge, such distinction has never been taken into account by any work to date. By looking at Fig.~\ref{fig.eps2_dist-relax} we observe that, for freestanding NCs, the relaxation clearly dominates over the total change of the absorption. This outcome allots to the structural strain a fundamental role in high-dense NC ensambles, and suggests the possibility of tuning the absorption characteristics by a controlled strain in light-driven applications, like for example photovoltaic cells or optical catalyzers. The possibility of performing strain relaxation in Si/SiO$_2$ NCs has been recently demonstrated by Arguirov et al.\cite{arguirov} 
using local laser annealing.


\subsection*{Orbitals Localisation and Interaction}\label{sec.localisation-interaction}
\begin{figure}[b!]
  \centering
  \includegraphics[draft=false,width=\columnwidth]{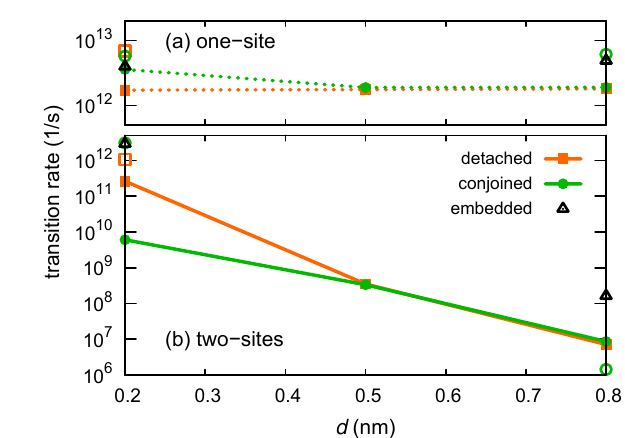}
  \caption{\small (color online) Optical transition rate for (A\,=\,Si$_{32}$, B\,=\,Si$_{17}$): two-sites HOMO$_A$--LUMO$_B$ transition (bottom panel) and one-site HOMO$_A$--LUMO$_A$ transition (top panel) as a function of the separation $d$. Squares and circles indicate values for detachedly and conjointly relaxed NC-NCs pairs, respectively. Filled and empty marks indicate freestanding and de-embedded NCs, respectively. Data for the embedded NCs is also reported by triangles. Lines are drawn to guide the eye.}\label{fig.matel_vs_d}
\end{figure}

\noindent Since in our system (made by two NCs) the absorption can be interpreted in terms of contributions from one-site and two-sites interband transitions (valence state to conduction state), it is not clear at this stage whether the modification of the absorption profile is governed by the former or by the latter transition type. In order to inquire into this aspect we have compared the optical transition rate (see Appendix A) of the one-site and two-sites interband transition of smallest energy. The characteristics of the transition, one- or two-sites, has been revealed by a plot of the involved orbitals.
\\In Fig.~\ref{fig.matel_vs_d} we report the HOMO$_A$--LUMO$_B$ (A\,=\,Si$_{32}$ and B\,=\,Si$_{17}$) transition rate $R$ as a function of $d$ and of the structural configuration. As expected, for freestanding detached and conjoined NCs, the one-site $R$ are stronger than the two sites ones for all the considered $d$. Moreover, for the detachedly or conjointly relaxed systems we observe a matching of the $R$ values at $d$\,=\,0.8\,nm and $d$\,=\,0.5\,nm both for one-site and two-sites transitions. Instead, at $d$\,=\,0.2\,nm a drastic reduction of $R$ (of a factor 50) appears in the case of conjoined relaxation for the HOMO$_A$--LUMO$_B$ two-site transition, while the one-site HOMO$_A$--LUMO$_A$ transition is enhanced of a factor of about two.
This result suggests that the two-sites HOMO$_A$--LUMO$_B$ transitions are much more sensitive to the strain-induced variations of the wave-function extent w.r.t.\ the one-site HOMO$_A$--LUMO$_A$. In fact, two-sites transitions require orbitals overlapping in the interstitial region between the NCs; in this picture and following the above result, the overlap increase at smaller $d$ due to orbital approaching, but decreases when the NCs are subjected to the strain. On the contrary, since the one-site $R$ depends on the overlap of orbitals localized on the same NC, it slightly decrease when the structures are rigidly approached (due to delocalization), while its increase during a conjoint relaxation indicates a re-localization of the orbitals.
Therefore, the total $R$ is explainable by a decrease of the orbital extent outside the NCs, enhancing the one-site oscillator strength to the detriment of the two-sites one. This result seems to suggests that, for freestanding small Si NCs, the structural rearrangement arising from NC-NC interaction at low $d$ tends to force the HOMO-LUMO delocalised states back into the NCs. In practice, the two-sites orbital overlap produce repulsive forces that screen the overlap itself.
\\The presence of negative values in Fig.~\ref{fig.eps2}b reveals that the effect of the strain over the states localization is not common to all the orbitals but is energy-dependent. As a consequence, the discussion above holds for interband transitions at energies approaching the band gap, but no generalization can be applied at this stage.
\\The case of embedded NCs is more complex and needs some additional arguments. In this case we expect that, beside inducing a severe strain on the NCs, the presence of a hosting matrix also limits the mobility of the system, reducing the possibility of structural rearrangements at low $d$. In addition, while at large $d$ the SiO$_2$ surrounding the NCs is able to partially compensate the stress emerging at the NCs interface, at small $d$ the few SiO$_2$ atoms in the interstitial NC-NC region cannot compensate the stress anymore, that is in this case maximized and shared between the NCs.
Therefore, the structural modifications at the origin of the modifications of the absorption profile of Fig.~\ref{fig.eps2_de-embedded} may not be produced by the NC-NC interaction forces, like in the freestanding case, but by the increase of the SiO$_2$-induced strain for the reason discussed above. The latter picture finds support in Fig.~\ref{fig.matel_vs_d} (see empty marks), where the $R$ at $d$\,=\,0.2\,nm of the de-embedded NCs shows a poor dependence on the relaxation type. In this case the trend of the one-site $R$ appears even inverted, indicating that the SiO$_2$-induced strain completely rules over the NC-NC mutual interaction forces and supporting the idea of a very limited capability of structural relaxations of the embedded NCs as function of the NC-NC separation.
\\In the case of embedded NCs we observe increased $R$ w.r.t.\ the freestanding counterparts. Also, the $R$ of de-embedded NCs match those of embedded ones, with the only exception of the two-sites $R$ at $d$\,=\,0.8\,nm that may be addressable to the different dielectric constants of SiO$_2$ and vacuum.
\\The severe strain induced by the embedding matrix seems to promote a delocalisation of the involved orbitals, since the two-sites $R$ is favoured at each $d$, while the one-site one seems to reduce at decreasing $d$. Moreover it is interesting to note that the delocalization is so strong in this case that at $d$\,=\,0.2\,nm the two-sites $R$ approaches the value of the one-site one. Following this picture we expect a strain-dependent NC-NC separation threshold for the formation of minibands in closely-packed NC arrays.\cite{seino}
\\The reduced possibility of embedded NCs to ``screen'' the presence of a neighbour NC renders the embedded systems an ideal candidate for applications that require strong energy/charge transfer between neighbouring NCs. At the opposite, in the case of freestanding, colloidal NC samples we expect a reduced NC-NC interplay (due to localized states).
\\We worth to stress that, following the superposition principle, in the limit case of non-interacting NCs the response of the ensamble shall be describable by the mere sum of the individual NC responses. Since in photoluminescence (PL) experiments the photo-generated exciton thermally dacays toward the band-edge before radiative recombination,\cite{PRB3} we suggest that the strain-induced variation of $R$ described above may be observable in PL experiments of colloidal samples by varying the NCs density.

\subsection*{Interband Optical Transitions}\label{sec.interband}
\noindent As anticipated above, the variation of the absorption with the strain is not constant but depends on the energy or, more specifically, on the initial and final transition states. To shed light on this point we must distinguish the optical transitions whose sum forms the absorption spectrum. To quantify the orbitals localization we make use of the inverse participation ratio (IPR), a numerical value connected to the ratio between the simulation box volume and the orbital volume (see Appendix B). To higher IPRs correspond higher orbital localizations and viceversa.
\begin{figure}[b!]
  \centering
  \includegraphics[draft=false,width=\columnwidth]{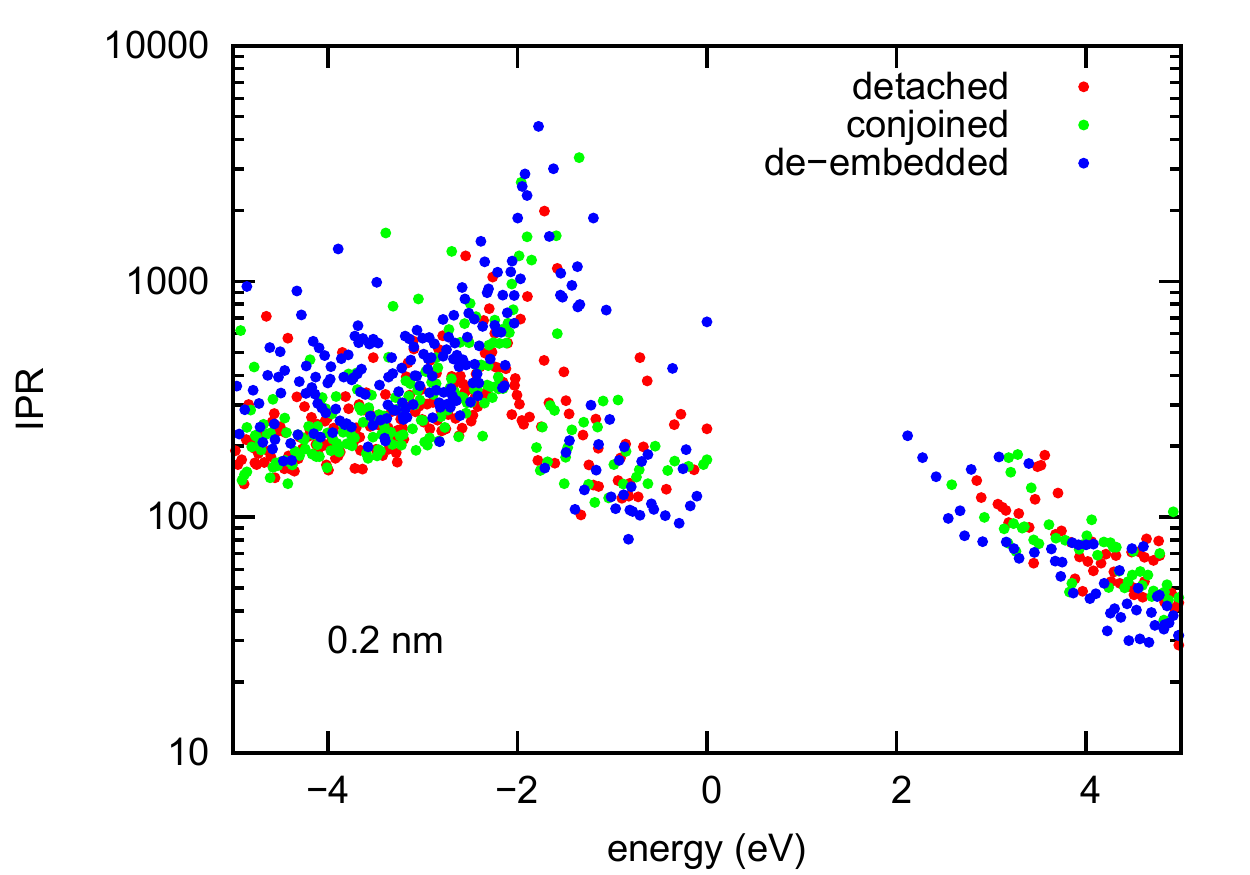}
  \caption{\small (color online) IPR values for the freestanding Si$_{32}$+Si$_{17}$ NCs pair placed at $d$\,=\,0.2\,nm as a function of the orbital energy and of the relaxation type. The zero of energy corresponds to the HOMO levels. }\label{fig.ipr}
\end{figure}
\begin{figure}[b!]
  \centering
  \includegraphics[draft=false,width=\columnwidth]{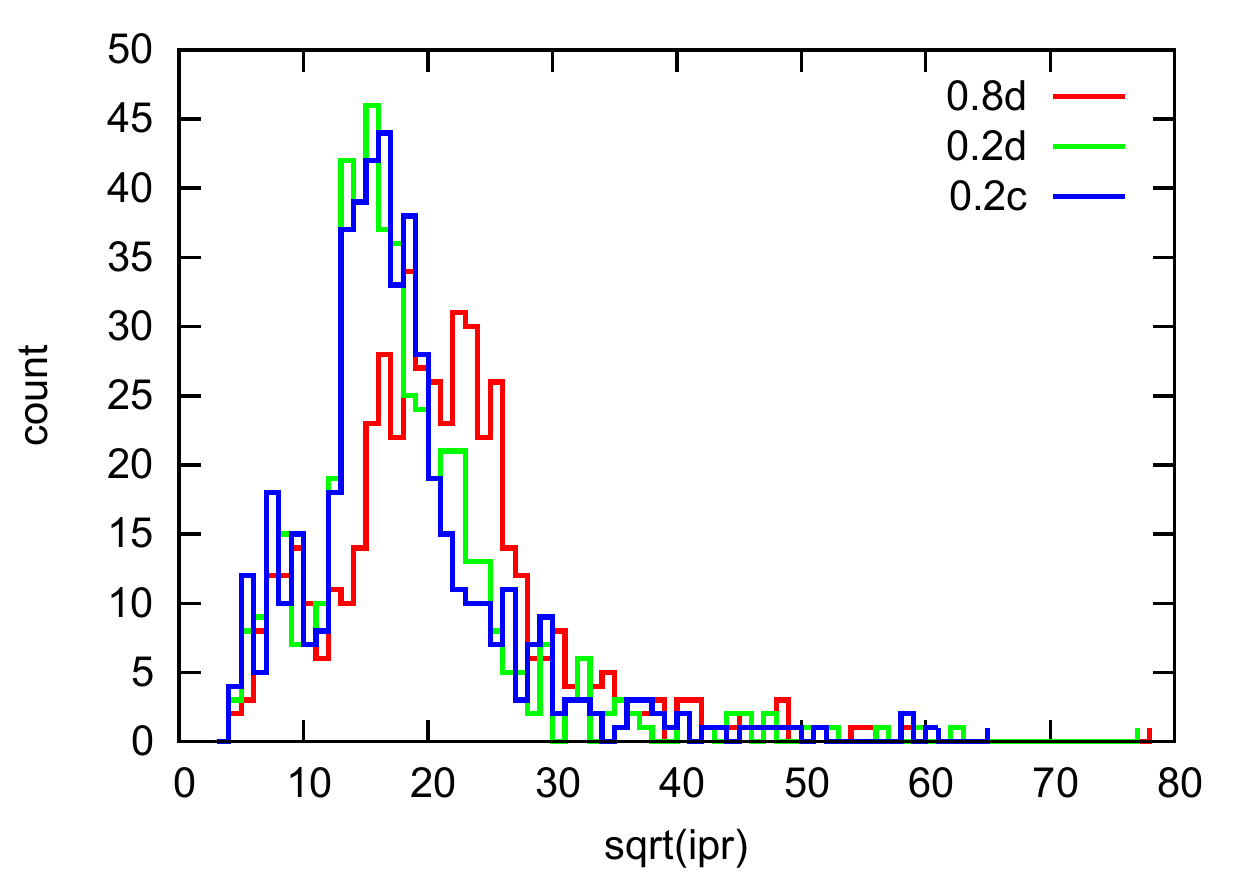}\\
  \includegraphics[draft=false,width=\columnwidth]{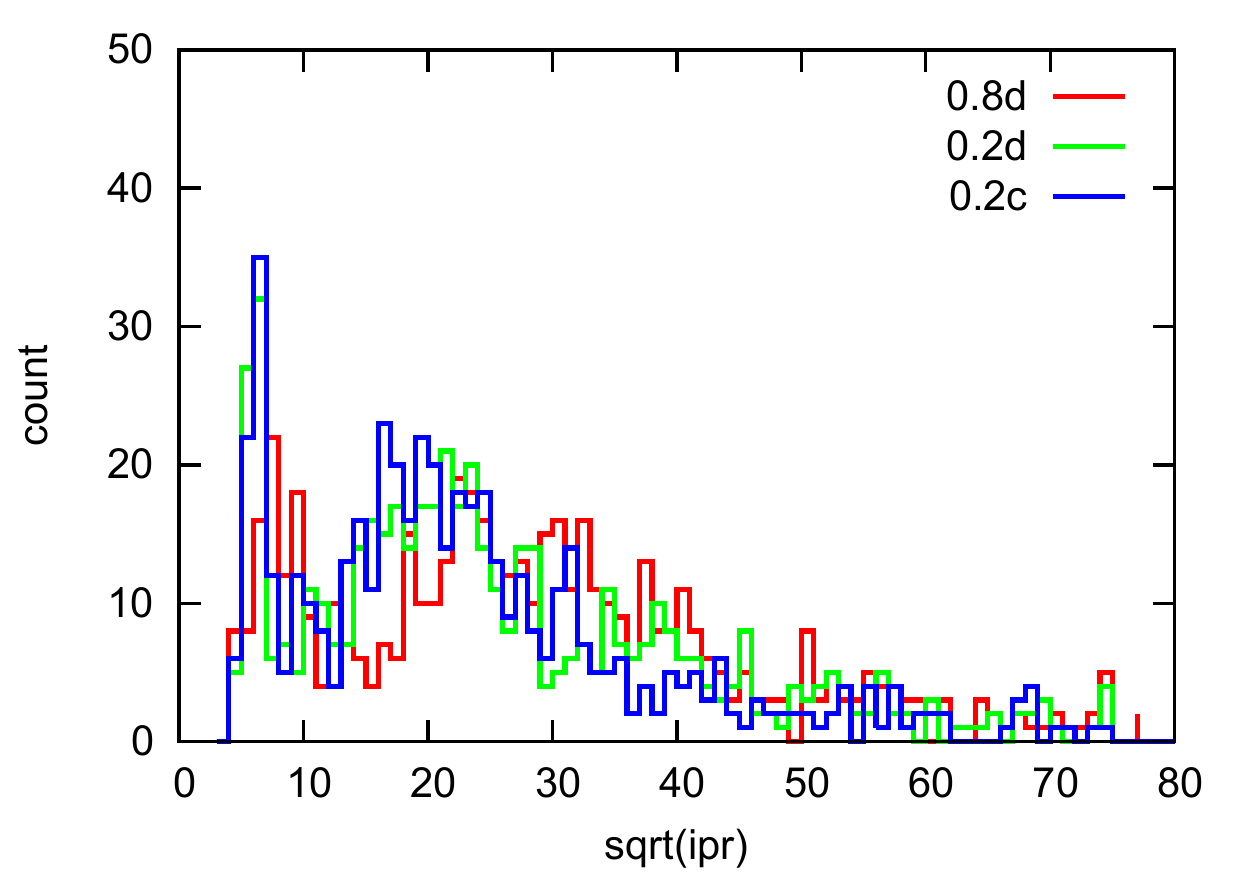}
  \caption{\small (color online) Distribution of the IPR values for the freestanding (top panel) or de-embedded (bottom panel) Si$_{32}$+Si$_{17}$ NCs pair. The curves report values for systems relaxed at $d$\,=\,0.8\,nm (0.8d), relaxed at $d$\,=\,0.8\,nm and then rigidly placed at $d$\,=\,0.2\,nm (0.2d), and conjointly relaxed at $d$\,=\,0.2\,nm (0.2c). }\label{fig.iprdistrib}
\end{figure}
\\In Fig.~\ref{fig.ipr} we report the IPRs for the Si$_{32}$+Si$_{17}$ NCs pair placed at $d$\,=\,0.2\,nm as a function of the orbital energy and of the relaxation type. Clearly, to an higher strain (de-embedded vs. conjoined vs detached) corresponds an increased spread between states with high and low IPRs: the IPR of deep valence states ($E$\,$\lesssim$\,-1.5\,eV) gets increased with the strain, while it is decreased for shallow valence and conduction states ($E$\,$\gtrsim$\,-1.5\,eV). The mutual NC-NC forces modify the NCs configuration during the conjoint relaxation, leading to an increased strain. However, the effect of the mutual NC-NC interaction is less pronounced w.r.t.\ that of an embedding SiO$_2$ matrix. The authenticity of the latter statement appears more clearly by comparing the IPR distribution of the NCs pair relaxed in vacuum (detached) with that relaxed in SiO$_2$ (de-embedded), as reported in Fig.~\ref{fig.iprdistrib}.
Clearly, independently on the NC-NC distance, in the latter case the low-IPRs peak ($\sqrt{IPR}$\,$\lesssim$\,12) increases of a factor of about two, the high-IPRs peak (12\,$\lesssim$\,$\sqrt{IPR}$\,$\lesssim$\,30) is reduced of about the same factor, while the number of very-high IPRs ($\sqrt{IPR}$\,$\gtrsim$\,30) slightly increase. Therefore, the overall effect of the strain on the IPRs appears as a broadening of the IPR distribution together with a boost of the low-IPRs peak and a damping of the high-IPRs one.
\\By comparing the 0.8d and 0.2d curves of Fig.~\ref{fig.iprdistrib} we also observe that, for a fixed NCs geometry, the sole reduction of the NC-NC distance entails a delocalisation of the orbitals (red vs. green curves). 
\\As discussed above, the variation of the orbitals localization with the distance/strain directly impacts the transition rate due to the connection between the matrix element and the initial-final states overlap (see Appendix A).
\begin{figure*}[[hbtp]
  \centering
  \includegraphics[draft=false,width=\textwidth]{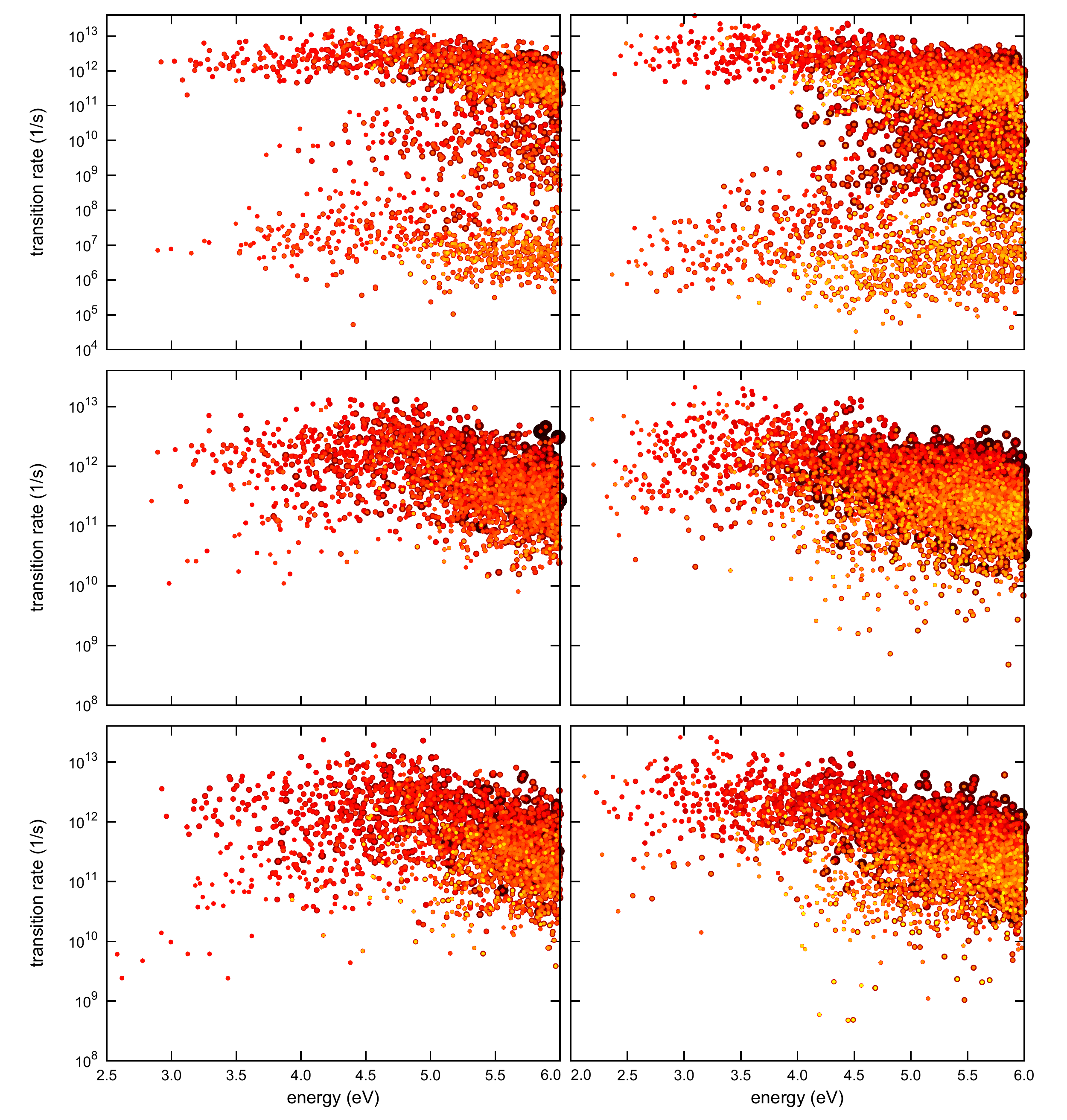}
  \caption{\small (colour online) Interband transition rates for the freestanding Si$_{32}$+Si$_{17}$ pair relaxed in vacuum (left panels) or in a SiO$_2$ matrix (de-embedded, right panels). Top panels report data for NCs relaxed at $d$\,=\,0.8\,nm, intermediate panels report values for NCs relaxed at $d$\,=\,0.8\,nm and then rigidly placed at $d$\,=\,0.2\,nm, and bottom panels report values for NCs relaxed at $d$\,=\,0.2\,nm. The IPR of the orbitals involved in each transition are indicated by the colour/size of the dots.
  Dark/large, red/medium, and yellow/small dots indicate states with IPR below, near to, and above the threshold value of 144 (see also Fig.~\ref{fig.iprdistrib}). For the sake of comparison the color/size scale of the dots is fixed for all the plots.}\label{fig.matel-ipr_multiplot}
\end{figure*}
In Fig.~\ref{fig.matel-ipr_multiplot} we report the transition rate vs. transition energy of the Si$_{32}$+Si$_{17}$ NCs pair relaxed in vacuum (left panels) or in a SiO$_2$ matrix (de-embedded, right panels) at $d$\,=\,0.8\,nm and $d$\,=\,0.2\,nm with different relaxation types. In the same Figure, the IPR of the orbitals associated to each transition rate is indicated by the color/size of the data points. At first we observe that at $d$\,=\,0.8\,nm the transition rates involving two states localised on different NCs (two-sites transitions, identifiable by a comparison with calculations for single NCs) are dramatically unfavoured and clearly separated by the one-site ones, especially at low energy. At higher energies the highly-delocalised valence orbitals come into play, forming a band of intermediate rates. Instead, the extremely-localised orbitals (deep states) produce high (one-site) or very-low (two-sites) rates because of the impossibility of overlapping in the interstitial region between the NCs.
\\We note that the transitions with the highest-rate have energies around 4.5\,eV for the system relaxed in vacuum, and around 3.5\,eV for the system relaxed in SiO$_2$. However, these energies don't correspond to the $\varepsilon_2$ maximum since the latter depends not only on the strength of the transitions but also on their number (see Eq.~\ref{eq.eps2-R} of Appendix A). Since the maximum achievable rate does not depend on $d$, it evidently pertains to one-site transitions. Besides, when the NCs are rigidly placed at $d$\,=\,0.2\,nm (Fig.~\ref{fig.matel-ipr_multiplot}, central panels) we observe that the two-sites $R$ approach the one-site ones, while a general orbital delocalisation appears, consistently with Fig.~\ref{fig.iprdistrib}. After the conjoint relaxation at $d$\,=\,0.2\,nm (Fig.~\ref{fig.matel-ipr_multiplot}, bottom panels) the re-localisation of the deep valence states (that are involved in high-energy transitions) is evidenced by an increased number of yellow dots in the Figure.
\\Consistently with Fig.~\ref{fig.matel_vs_d}, in the case of freestanding NCs the conjoint relaxation boosts the one-site $R$ while reducing the two-sites ones. Instead, as discussed above (see Fig.~\ref{fig.iprdistrib}), in the case of de-embedded NCs the high SiO$_2$-driven strain forms a large number of extremely-localised states in a background of very-delocalised states. The large strain makes difficult to recognize changes of $R$ driven by the NC-NC interaction forces.
\\Finally we note that, since the strain can either increase or reduce the IPR depending on the orbital energy, such dependency on the energy is reflected on the transition rates and finally on the $\varepsilon_2$. Therefore the applied strain produces higher or lower rates depending on the energy of the initial and final state of the transition. The above considerations explain the alternating variation of the absorption shown in Fig.~\ref{fig.eps2}b.

\subsection*{Intraband Optical Transitions}\label{sec.intraband}
\begin{figure}[t!]
  \centering
  \includegraphics[draft=false,width=\columnwidth]{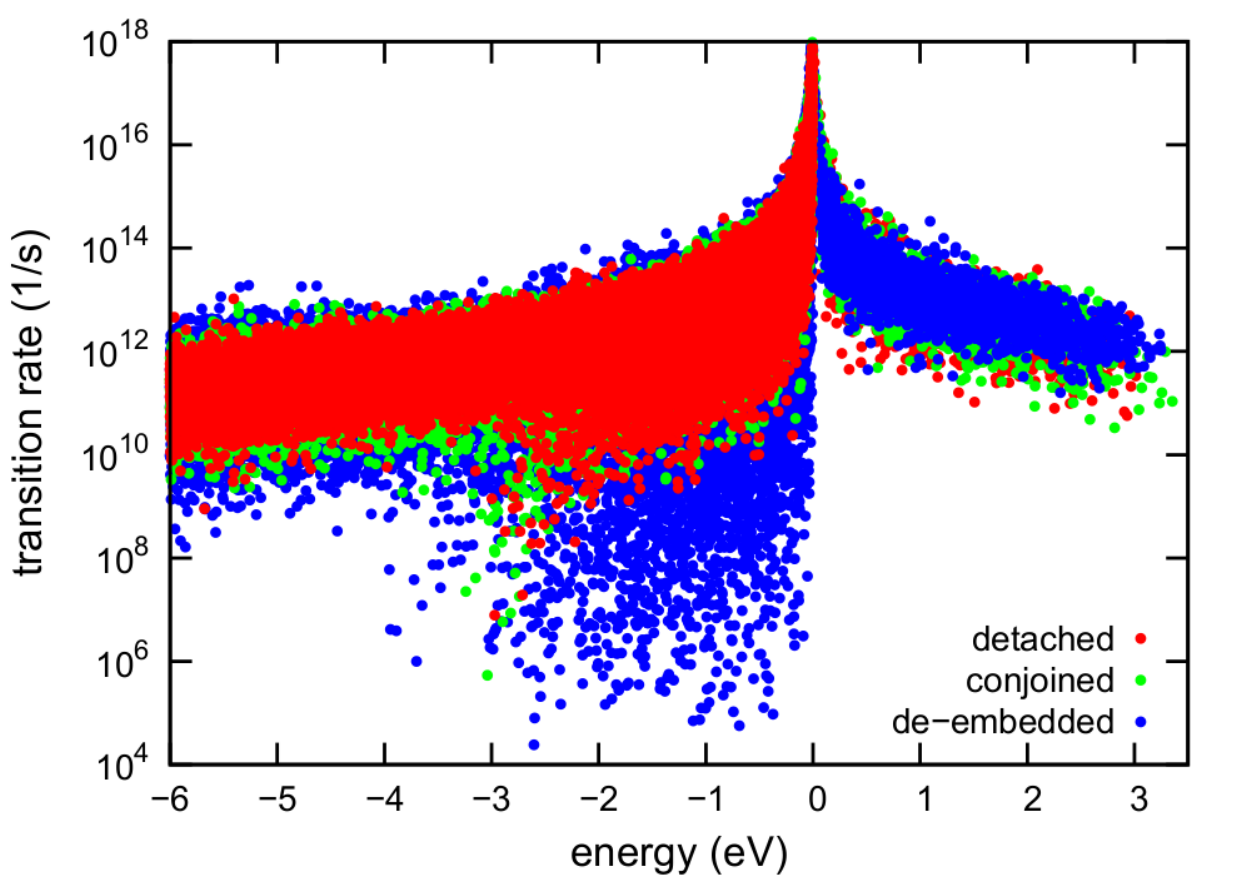}
  \caption{\small (color online) Intraband transition rates for the Si$_{32}$+Si$_{17}$ NCs pair relaxed in vacuum at $d$\,=\,0.8\,nm and then rigidly placed at $d$\,=\,0.2\,nm (detached), conjointly relaxed at $d$\,=\,0.2\,nm (conjoined), and relaxed in SiO$_2$ at 0.8\,nm and then de-embedded and rigidly placed at $d$\,=\,0.2\,nm (de-embedded). Positive and negative energies refer to transitions in the conduction and valence band, respectively.}\label{fig.intraband}
\end{figure}
\noindent While interband transitions are essentially related to optical absorption/emission, intraband transitions plays a fundamental role in processes such as SSQC\cite{timmerman,trinh,ivan}, 
MEG,\cite{ivan,timmerman2}, transport,\cite{balberg} and others. The possibility of controlling the rate of the above processes in dense NC ensambles have attracted a lot of interest in the latest years \cite{ivan}. Besides, the role of strain on the above mechanisms has been often neglected. It is therefore important to understand how the above results apply in the case of intraband transitions.
\\Intraband transitions can be performed by electrons transiting on the conduction band, or by holes transiting on the valence band. Since strain have opposite effects on orbitals belonging to deep valence and shallow valence and conduction band, we expect an asymmetric effect on the transition rates of electron and holes .
\\As shown in Fig.~\ref{fig.ipr}, in the case of conduction states the strain uniformly delocalise the orbitals, favouring the coupling between states belonging to different NCs. Therefore, we expect in this case an enhanced possibility of energy- or electron-transfer between neighbouring NCs at higher strain levels.
\\In the case of valence states one has to distinguish between shallow states and deep states. Since IPR changes are opposite for shallow and deep valence states, we expect a large effect of the strain on the transition rates.
\\The scenario outlined above emerges in Fig.~\ref{fig.intraband} in which the intraband transition rates for electrons (positive energies) and for holes (negative energies) are presented for systems subject to different strain levels. Clearly, transitions in the valence band are dramatically interfered by the strain, while transitions in the conduction band are unaltered or possibly favoured, especially at high energy. This result may explain the dominance of electron transport over hole transport in Si/SiO$_2$ embedded NCs,\cite{seino} and strongly fosters the employment of strained structures in future experiments\cite{kusova} to verify the possibility of enhancing or reducing the NC-NC interaction mechanisms in samples with high NCs density.

\section{Conclusions}\label{sec.conclusions}
In dense NC ensambles, the overlap of the wavefunctions of two neighbouring NCs produces a reconfiguration of the electronic structure of the NCs, and a structural rearrangement following the mutual NC-NC interaction forces. In agreement with other works,\cite{timmerman,trinh,ivan,galli,seino} our results indicate that a close packing of Si-NCs is required in order to evidence any kind of interaction effect. For our systems we have measured a threshold surface-to-surface NC-NC distance of about 0.5\,nm, below which we observe some kind of electronic and ionic structural reconfiguration.
\\By relaxing the NCs in vacuum or in a SiO$_2$ matrix and by rigidly placing the NCs in the simulation box or by permitting a full conjoint relaxation of the NCs pair including the mutual NC-NC interaction forces, we have been able to distinguish the effect of the NC-NC distance by that of the strain on the optical and electronic properties of the system.
\\At first we have observed that the sole reduction of $d$ produces a general increase of the orbitals extent, promoting the rate of two-sites transitions. Besides, the sole increase of the strain acts differently on conduction states, on shallow valence states, and on deep valence states. While the former two gets delocalised, the latter increase their localization, producing a complex response that strongly depends on the transition energy.
\\In the case of optical absorption (i.e. interband transitions) we have shown that for freestanding NCs the structural rearrangement induced by the mutual NC-NC interaction forces dominates the spectral modifications w.r.t.\ the sole variation of the distance.
\\In the presence of an embedding matrix the effect of the mutual NC-NC interaction forces is reduced, while large matrix-induced strain maximizes the orbital localization/delocalisation effect.
\\In the case of intraband transitions, we have shown that transitions within the conduction band are promoted by higher strain levels, while transitions within the valence band are strongly limited. The latter result may explain the dominance of electrons over holes in the transport properties of SiO$_2$-embedded Si NCs.\cite{seino}
\\More in general, our results reveal that for neighbouring NCs the strain may have a great influence over any kind of NC-NC interaction mechanisms, suggesting the possibility of a strain-enabled and process-dependent control of the response in colloidal\cite{fujii} or embedded\cite{arguirov} Si-NC ensambles.

\vspace{0.5cm}
\noindent {\footnotesize{\bf ACKNOWLEDGEMENTS:} Computational resources were made available by CINECA-ISCRA parallel computing initiative. We acknowledge financial support from the European Community's Seventh Framework Programme (FP7/2007-2013) under Grant No. 245977.}

\section*{APPENDIX A: Transition Rate}\label{sec.appendix_rate}
\noindent The rate of absorption or emission processes of radiation with energy $\hbar\omega$, intensity $A_0$, and wavevector $\textbf{q}$ is defined from the Fermi golden rule and is given by\cite{grosso}
\begin{equation}
W(\textbf{q},\omega) = \frac{2 \pi}{\hbar}\,2\,\sum_{i,j}|\langle j | A_0e^{i\textbf{q}\cdot\textbf{r}} | i \rangle |^2\delta(E_{ij}-\hbar\omega)[f_i-f_j]~, \label{eq.trans_rate}
\end{equation}
where $E_{ij}$\,=\,$E_i$\,--\,$E_j$, $f_i$ and $f_j$ are the occupations of states $|i\rangle$ and $|j\rangle$, and the factor 2 in front of the summation takes into account the spin degeneracy. In the case of absorption the sum is performed over valence states $|i\rangle$ with $f_i$\,=\,1, and conduction states $|j\rangle$ with $f_j$\,=\,0, with $E_{ij}>0$.
The imaginary part of the dielectric function, $\varepsilon_2$ is related to $W$ by
\begin{equation}
\varepsilon_2(\textbf{q},\omega) = \frac{2\pi\hbar e^2}{q^2}\frac{1}{V}\frac{W(\textbf{q},\omega)}{|A_0|^2}~, \label{eq.eps2-W}
\end{equation}
where $e$ is the electron charge and $V$ is the cell volume. In the long wavelength limit we have
\begin{equation}
\langle j | e^{i\textbf{q}\cdot\textbf{r}} | i \rangle = i\textbf{q}\langle j | \textbf{r} | i \rangle = \frac{\hbar}{m}\frac{i\textbf{q}\cdot \langle j | \textbf{p} | i \rangle}{E_{ij}}~, \label{eq.dip-approx}
\end{equation}
$m$ being the rest mass of electron and $\textbf{p}$ the momentum operator. From Eq.~\ref{eq.trans_rate}-\ref{eq.dip-approx} we can express the contribution to $\varepsilon_2$ of each interband transition $i\rightarrow j$, simply referred here as the \textit{transition rate}
\begin{equation}
R_{ij} = \frac{8\pi^2e^2\hbar^2}{m^2V}\frac{|\langle j|\textbf{p}|i\rangle|^2}{E_{ij}^2}~. \label{eq.R}
\end{equation}
Then, $\varepsilon_2$ is related to $R$ by
\begin{equation}
\varepsilon_2(\omega) = \sum_{i,j}R_{ij}\delta(E_{ij}-\hbar\omega)~. \label{eq.eps2-R}
\end{equation}
Note that $R$ and $\varepsilon_2$ depend on the cell volume $V$. For all the calculations we used a cubic simulation box with side 2.6\,nm, giving $V$=17.58\,nm$^3$.

\section*{APPENDIX B: Inverse Participation Ratio }\label{sec.appendix_IPR}
\noindent For a given wavefunction $\psi_n$ the Inverse Participation Ratio (IPR) is defined as:
\begin{equation}
IPR(\psi_n)=V\frac{ \int\limits_V \left| \psi_n(\vec{r}) \right|^4 }{ \left[ \int\limits_V \left| \psi_n(\vec{r}) \right|^2 \right]^2 } , \label{eq.ipr0}
\end{equation}
where $V$ is the volume of the simulation box. Eq. \ref{eq.ipr0} returns unity for a maximally dispersed state and infinity for a maximally localized state. In numerical simulations the cell is divided into a finite grid, and Eq. \ref{eq.ipr0} reduces to:
\begin{equation}
IPR(\psi_n)=N\frac{ \sum\limits_{i=1}^{N} \left| \psi_n(\vec{r}_i) \right|^4 }{ \left[ \sum\limits_{i=1}^{N} \left| \psi_n(\vec{r}_i) \right|^2 \right]^2 } , \label{eq.ipr}
\end{equation}
where the sum is performed over the $N$ volume elements of the grid. Thus the IPR is limited to $N$ in the case of a maximally localized state. Simply speaking, the IPR is connected to the ratio between the cell volume and the orbital volume. As a reference, Eq.\,\ref{eq.ipr} applied on an hypotetical orbital homogeneously occupying a volume of a sphere with diameter of 1\,nm in our calculations would return IPR\,$\simeq$\,33. Instead, the same orbital distributed gaussianly with a full-width at tenth-maximum of 1\,nm would have IPR\,$\simeq$\,40. In all our calculation we used a 253$\times$253$\times$253 uniform volume grid in the real space, giving $N$\,=\,253$^3$.


\footnotesize
\begin{thebibliography}{}
\bibitem{nature_pavesi} L. Pavesi, L. Dal Negro, C. Mazzoleni, G. Franz\`o, F. Priolo, Nature {\bf 408}, 440 (2000).
\bibitem{bourianoff} G. I. Bourianoff and  H. A. Atwater, Nature Materials {\bf 4}, 143 (2005).
\bibitem{tiwari} S. Tiwari, F. Rana, H. Hanafi, A. Hartstein, E. F. Crabbe, K. Chan, Appl. Phys. Lett. {\bf 68} (1996).
\bibitem{yao} J. Yao, Z. Sun, L. Zhong, D. Natelson, J. M. Tour, Nano Lett. {\bf 10}, 4105 (2010).
\bibitem{erogbogbo} F. Erogbogbo, K-T Yong, I. Roy, R. Hu, W-C Law, W. Zhao, H. Ding, F. Wu, R. Kumar, M. T. Swihart, and P. N. Prasad, ACS Nano {\bf 5}, 413 (2011).
\bibitem{osso_nanoreslett} S Ossicini, M. Amato, R. Guerra, M. Palummo, O. Pulci, Nanoscale Res. Lett. {\bf 5}, 1637 (2010).
\bibitem{heitmann_zacharias} J. Heitmann, F. M\"{u}ller, L. Yi, M. Zacharias, D. Kovalev, F. Eichhorn, Phys. Rev. B {\bf 69}, 195309 (2004).
\bibitem{linnros} J. Linnros, N. Lalic, A. Galeckas, v. Grivickas, J. Appl. Phys. {\bf 86}, 6128 (1999).
\bibitem{glover_meldrum} M. Glover, A. Meldrum, Optical Materials {\bf 27}, 977 (2005).
\bibitem{iwayama} T. Shimitsu-Iwayama, T. Hama, D. E. Hole, I. W. Boyd, Solid-State Electronics {\bf 45}, 1487 (2001).
\bibitem{schneibner} M. Schneibner, M. Yakes, A. S. Bracker, I. V. Ponomarev, M. F. Doty, C. S. Hellberg, L. J. Whitman, T. L. Reinecke, D. Gammon, Nature Physics {\bf 4}, 291 (2008).
\bibitem{timmerman} D. Timmerman, I. Izeddin, P. Stallinga, I. N. Yassievich, T. Gregorkiewicz, Nature Photonics {\bf 2}, 105 (2008).
\bibitem{trinh} M. T. Trinh, R. Limpens, W. D. A. M. de Boer, J. M. Schins,
L. D. A. Siebbeles, T. Gregorkiewicz, Nature Photonics {\bf 6}, 316 (2012).
\bibitem{ivan} M. Govoni, I. Marri, S. Ossicini, Nature Photonics {\bf 6}, 672 (2012).
\bibitem{ossicini} S. Ossicini, L. Pavesi, F. Priolo, {\it Light Emitting Silicon for Microphotonics}, Springer Tracts on Modern Physics 194 (Springer-Verlag, Berlin,2003).
\bibitem{degoli} E. Degoli, S. Ossicini, Adv. Quantum Chem. {\bf 58},  203 (2009).
\bibitem{guerrasin} R. Guerra, M.  Ippolito, S. Meloni, S. Ossicini, Appl. Phys. Lett. {\bf 100}, 181095 (2012).
\bibitem{dalnegro} L. Dal Negro, J. H. Yi, L. C. Kimerling, S. Hamel, A. Williamson,, G. Galli Appl. Phys. Lett. {\bf 88}, 183103 (2006) 
\bibitem{loeper} P. L\"{o}per, R. M\"{u}ller, D. Hiller, T. barthel, E. malguth, S. Janz, J. C. Goldschmidt, M. Hermle, M. Zacharias, Phys. Rev. B {\bf 84}, 195317 (2011).
\bibitem{zacharias} M. Zacharias, J. Heitmann, R. Scholz, U. Kahler, M. Schmidt, J. Bl\"{a}sing, Appl. Phys. Lett. {\bf 80}, 661 (2002).
\bibitem{lockwood_meldrum} R. Lockwood, A. Hryciw, A. Meldrum, Appl. Phys. Lett. {\bf 89}, 263112 (2006).
\bibitem{godefroo} S. Godefroo, M. Hayne, M. Jivanescu, A. Stesmans, M. Zacharias, O. I. Lebedev, G. Van Tendeloo, V. V. Moshchalkov, Nature Nano {\bf 3}, 174 (2008).
\bibitem{allan_delerue} G. Allan, C. Delerue, Phys. Rev. B {\bf 75}, 195311 (2007).
\bibitem{galli} A. Gali, M. V\"{o}r\"{o}s, D. Rocca, G. T. Zimanyi, G. Galli, Nano Letters {\bf 9}, 3780 (2009).
\bibitem{seino} K. Seino, F. Bechstedt, P. Kroll, Phys. Rev. B {\bf 86}, 075312 (2012).
\bibitem{lusk} Z. Lin, H. Li, A. Franceschetti, M. T. Lusk, ACS Nano {\bf 6}, 4029 (2012).
\bibitem{natmat_osso} M. Cazzanelli, F. Bianco, E. Borga, G. Pucker, M. Ghulinyan, E. Degoli, E. Luppi, V. V\'eniard, S. Ossicini, D. Modotto, S. Wabnitz, R. Pierobon,  L. Pavesi, Nature Materials {\bf 11}, 148 (2012).
\bibitem{nduwimana} A. Nduwimana and X-Q Wang, J. Phys. Chem. C {\bf 114}, 9702 (2010). 
\bibitem{beard} M. C. Beard, K. P. Knutsen, P. Yu, J. M. Luther, Q. Song,W. K. Metzger, R. J. Ellingson, A. J. Nozik, Nano Lett. {\bf 7}, 2506 (2007).
\bibitem{fujii} H. Sugimoto, M. Fujii, K. Imakita, S. Hayashi, K. Akamatsu, J. Phys. Chem. C {\bf 116}, 17 969 (2012).
\bibitem{PRB1} R. Guerra, I. Marri, R. Magri, L. Martin-Samos, O. Pulci, E. Degoli, S. Ossicini, Phys. Rev. B {\bf 79}, 155320 (2009).
\bibitem{PRB2} R. Guerra, E. Degoli, S. Ossicini, Phys. Rev. B {\bf 80}, 155332 (2009).
\bibitem{espresso} P. Giannozzi et al., J. Phys. Condens. Matter {\bf 21}, 395502 (2009).
\bibitem{onida_RevModPhys} G. Onida, L. Reining, A. Rubio, Rev. Mod. Phys. {\bf 74}, 601 (2002).
\bibitem{makov-payne} G. Makov, M. C. Payne, Phys. Rev. B {\bf 51}, 4014 (1995).
\bibitem{kusova} K. K\r{u}sov\'a, L. Ondi\v{c}, E. Klime\v{s}ov\'a, K. Herynkov\'a, I. Pelant, S. Dani\v{s}, J. Valenta, M. Gallart, M. Ziegler, B. Ho\"nerlage, P. Gilliot, Appl. Phys. Lett. {\bf 101}, 143101 (2012).
\bibitem{arguirov} T. Arguirov, T. Mchedlidze, M. Kittler, R. R\"olver, B. Berghoff, M. F\"orst, B. Spangenberg, Appl. Phys. Lett. {\bf 89}, 053111 (2006).
\bibitem{PRB3} R. Guerra, S. Ossicini, Phys. Rev. B {\bf 81}, 245307 (2010).
\bibitem{timmerman2} D. Timmerman, J. Valenta, k. Dohnalova, W. D. A. M. de Boer, and T. Gregorkiewicz, Nature Nanotech. {\bf 6}, 710 (2011).
\bibitem{balberg} I. Balberg, E. Savir, J. Jedrzejewski, A. G. Nassiopoulou, S. Gardelis, Phys. Rev. B {\bf 75}, 235329 (2007).
\bibitem{grosso} G. Grosso and G. P. Parravicini, in \textit{Solid State Physics}, Academic Press 2003, ISBN 0-12-304460-X.

\end{thebibliography}
\end{document}